\documentclass[twocolumn,showpacs,preprintnumbers,amsmath,amssymb,groupedaddress,superscriptaddress]{revtex4}

\usepackage[dvips]{graphicx}
\usepackage[dvips]{color}

\usepackage{dcolumn}
\usepackage{bm}
\newcommand{\ignore}[1]{}

\begin{document}
\def\e{\mathcal{E}}

\title{Optimal control of light pulse storage and retrieval}
\author{Irina Novikova}
\affiliation{Harvard-Smithsonian Center for Astrophysics, Cambridge,
Massachusetts 02138, USA}\affiliation{Department of Physics, College of
William\&Mary, Williamsburg, Virginia 23185, USA}
\author{Alexey V. Gorshkov}
\affiliation{Department of Physics, Harvard University, Cambridge,
Massachusetts 02138, USA}
\author{David F. Phillips}
\affiliation{Harvard-Smithsonian Center for Astrophysics, Cambridge,
Massachusetts 02138, USA}
\author{Anders S. S{\o}rensen}
\affiliation{QUANTOP, Danish National Research Foundation Centre of Quantum
Optics, Niels Bohr Institute, DK-2100 Copenhagen {\O}, Denmark}
\author{Mikhail D. Lukin}
\affiliation{Department of Physics, Harvard University, Cambridge,
Massachusetts 02138, USA}
\author{Ronald L. Walsworth}
\affiliation{Harvard-Smithsonian Center for Astrophysics, Cambridge,
Massachusetts 02138, USA} \affiliation{Department of Physics, Harvard
University, Cambridge, Massachusetts 02138, USA}

\date{\today}

\begin{abstract}
We demonstrate experimentally a procedure to obtain the maximum efficiency for
the storage and retrieval of light pulses in atomic media. The procedure uses
time reversal to obtain optimal input signal pulse-shapes. Experimental results
in warm Rb vapor are in good agreement with theoretical predictions and
demonstrate a substantial improvement of efficiency. This optimization
procedure is applicable to a wide range of systems.

\end{abstract}


%
%

\maketitle

Mapping of quantum states between light and matter is a topic of great current
interest~\cite{lukin03rmp,julsgaard04,kraus06}.  One of the leading approaches
to realizing this capability is the storage of light in ensembles of radiators
(warm atoms, cold atoms, impurities in solids, etc.) using a dynamic form of
electromagnetically induced transparency
(EIT)~\cite{harrispt,fleischhauerrmp,fleischhauer,liu,phillips,hemmer,lukin03rmp}.
This light storage technique has been shown experimentally to preserve optical
phase coherence~\cite{mair} as well as quantum correlations and
states~\cite{kuz03etal,vdwal03}; and thus has emerged as a promising technique
for applications such as single-photon generation on
demand~\cite{chou04,matsukevich04,eisaman04,matsukevich06,chen06} and quantum
memories~\cite{chou05,kuzmich05,eisaman05} and
repeaters~\cite{DLCZ,kuzmich06,felinto06}.  However, practical applications
will require significant improvements in the efficiency of writing, storing and
retrieving an input photon state beyond values achieved to date
~\cite{novikova05,felinto05,laurat06,thompson06}. As an advance in this
direction, we report in this Letter an experimental demonstration 
of an optimization protocol based on time reversal~\cite{gorshkov}, which
determines the input signal pulse-shape that is written, stored, and retrieved
with maximum efficiency for a given set of experimental conditions. This
optimization procedure should be applicable to a wide range of ensemble systems
in both classical and quantum regimes.

\begin{figure}
\includegraphics[width=\columnwidth]{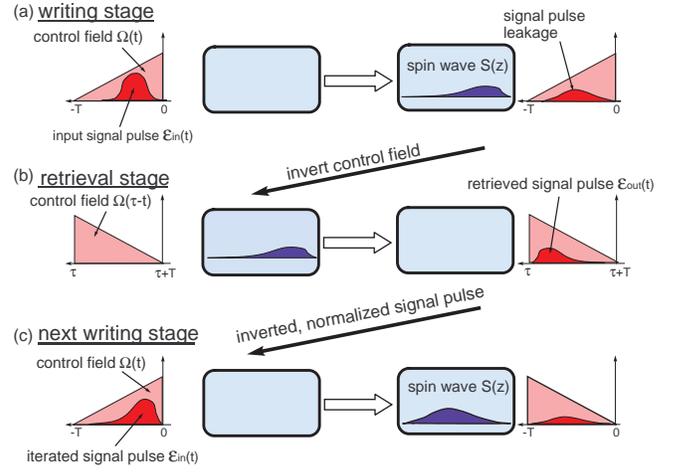}
\caption{(color online) Schematic of the iterative signal pulse optimization
procedure described in the text. (a) An input signal pulse $\e_\textrm{in}(t)$
is mapped into a spin-wave $S(z)$ using a control field envelope $\Omega(t)$.
(b) After a storage period $\tau$, the spin-wave is mapped into an output
signal pulse $\e_\textrm{out}$ using the time-reversed control field envelope
$\Omega(\tau-t)$. (c) The time-reversed and normalized version  of the measured
$\e_ \textrm{out}$ is used as the input $\e_\textrm{in}$ in the next iteration.
} \label{cartoon.fig}
\end{figure}

We consider the interaction of a weak (classical or quantum) signal pulse and a
strong (classical) control field with a $\Lambda$-type medium. While the
optimization procedure is applicable to a wide class of regimes,  we focus here
on resonant light-atom interactions under EIT conditions. The group velocity of
the signal pulse is proportional to the intensity of the control field, such
that the quantum state of the signal pulse can be reversibly stored in a
collective spin coherence (spin-wave) of the atomic ensemble by reducing the
control field to zero~\cite{lukin03rmp}.  For ideal writing, storage, and
retrieval, the signal pulse's frequency components must fit well within the EIT
spectral window ($\Delta \omega_\textrm{EIT}$) to avoid incoherent absorptive
loss: i.e., $1/t_s \ll \Delta \omega_\textrm{EIT} \simeq \sqrt{d}
v_\textrm{g}/L$~\cite{fleischhauer}, where $t_s$ is the temporal length of the
signal pulse, $v_\textrm{g}$ is the group velocity of the signal pulse inside
the EIT medium at full control field intensity, $L$ is the length of the
medium, and $d$ is the optical depth of the medium for the signal pulse in the
absence of EIT conditions~\cite{dnote}. In addition, $v_\textrm{g}$ must be
small enough for the entire signal pulse to be spatially compressed into the
ensemble before storage --- i.e., $v_\textrm{g}t_s \ll L$
--- so as to avoid ``leakage'' of the front edge of the pulse outside the
medium before the back edge has entered. Simultaneous satisfaction of both
these conditions is possible only at very large optical depth $d$, i.e., at
high density and/or large sample size. However, operation at very large $d$ can
degrade EIT performance and shorten the spin-wave coherence lifetime due to
radiation trapping, competing nonlinear processes, etc.  Therefore, practical,
high-efficiency light storage will likely be performed at moderately large $d$
and require optimization of the input signal pulse-shape to minimize absorptive
and leakage losses.

Recently, a procedure to determine the optimal input signal pulse-shape for a
given optical depth and control field was proposed~\cite{gorshkov}.
This  optimization procedure is based on successive time-reversal iterations
and shown schematically in Fig.~\ref{cartoon.fig}.  First, for a given input
control field with Rabi frequency envelope $\Omega(t)$, a trial input signal
pulse with envelope $\e_\textrm{in}(t)$ is mapped into a spin-wave $S(z)$
inside the atomic ensemble (writing stage). ($\e_\textrm{in}(t)$ and the input
control field are taken to be non-zero over the time-interval $[-T,0]$.) In
general there will be some absorptive and leakage losses during this writing
process. After a storage period $\tau$, an output control field
$\Omega(\tau-t)$ --- i.e., the time-reversed version of the input control field
--- is used to map $S(z)$ back into an output signal pulse
$\e_\textrm{out}(t)$, which leaves the medium and is measured (retrieval
stage). The input signal pulse for the next iteration is then generated with a
pulse-shape corresponding to a time-reversed version of the previous output
signal pulse and an amplitude normalized to make the energy of the pulse equal
to a fixed target value. These steps are then repeated iteratively, using the
same input and output control fields, until the shape of the output signal
pulse on a given iteration is identical to the time-reversed profile of its
corresponding input signal pulse. The resulting signal pulse-shape is predicted
to provide the highest write/store/retrieve efficiency possible for a given
optical depth and control field profile, and should be applicable to both
quantum and weak classical signal pulses.

\begin{figure*}
\includegraphics[width=1.3\columnwidth]{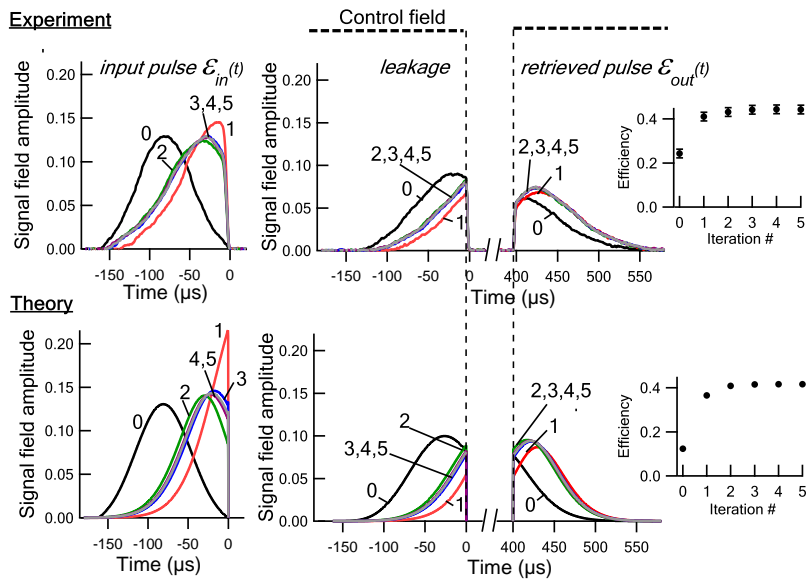}
\caption{(color online) \underline{Top}: Example data for the signal pulse
optimization procedure, using a constant control field during writing and
retrieval (timing indicated by dashed lines) and a $400~\mu$s storage interval.
\textit{Left:} input signal pulses $\e_\textrm{in}$, labeled by the iteration
number and beginning with a trial Gaussian input pulse (iteration ``0'').
\textit{Center:} signal pulse leakage for each iteration. \textit{Right:}
output signal pulse $\e_\textrm{out}$ for each iteration. \textit{Inset}:
Write/store/retrieve efficiency determined for each iteration from the measured
input and output signal pulses: $\int \e^2_\textrm{out}dt/\int
\e^2_\textrm{in}dt$. \underline{Bottom}: theoretical calculation of the signal
pulse optimization procedure, using the model described in the text and the
experimental conditions of the measurements in the top panel. \textit{Inset}:
Calculated write/store/retrieve efficiency. } \label{SampleProbePulses.fig}
\end{figure*}

In the experiment reported here, we tested this optimization procedure and
confirmed its three primary predictions:

\emph{1.} The write/store/retrieve efficiency (the ratio of energies carried by
the retrieved and input signal pulses) grows with each iteration until the
input signal field converges to an optimal pulse-shape. See
Fig.~\ref{SampleProbePulses.fig}.

\begin{figure}
\includegraphics[width=1.0\columnwidth]{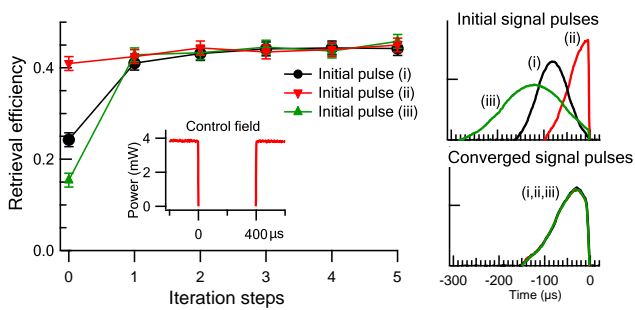}
\caption{(color online) Example data illustrating that the iterative
optimization procedure converges to the same signal pulse-shape and maximum
efficiency independent of the initial signal pulse-shape, for a given control
field profile (shown in inset) and optical depth ($d$=9.0). }
\label{VarSigPulses.fig}
\end{figure}

\emph{2.} For a given control field profile and optical depth $d$, the
optimization procedure converges to the same input signal pulse-shape and the
same maximum efficiency, independent of the initial (trial) signal pulse-shape.
See Fig.~\ref{VarSigPulses.fig}.

\begin{figure*}
\includegraphics[width=1.5\columnwidth]{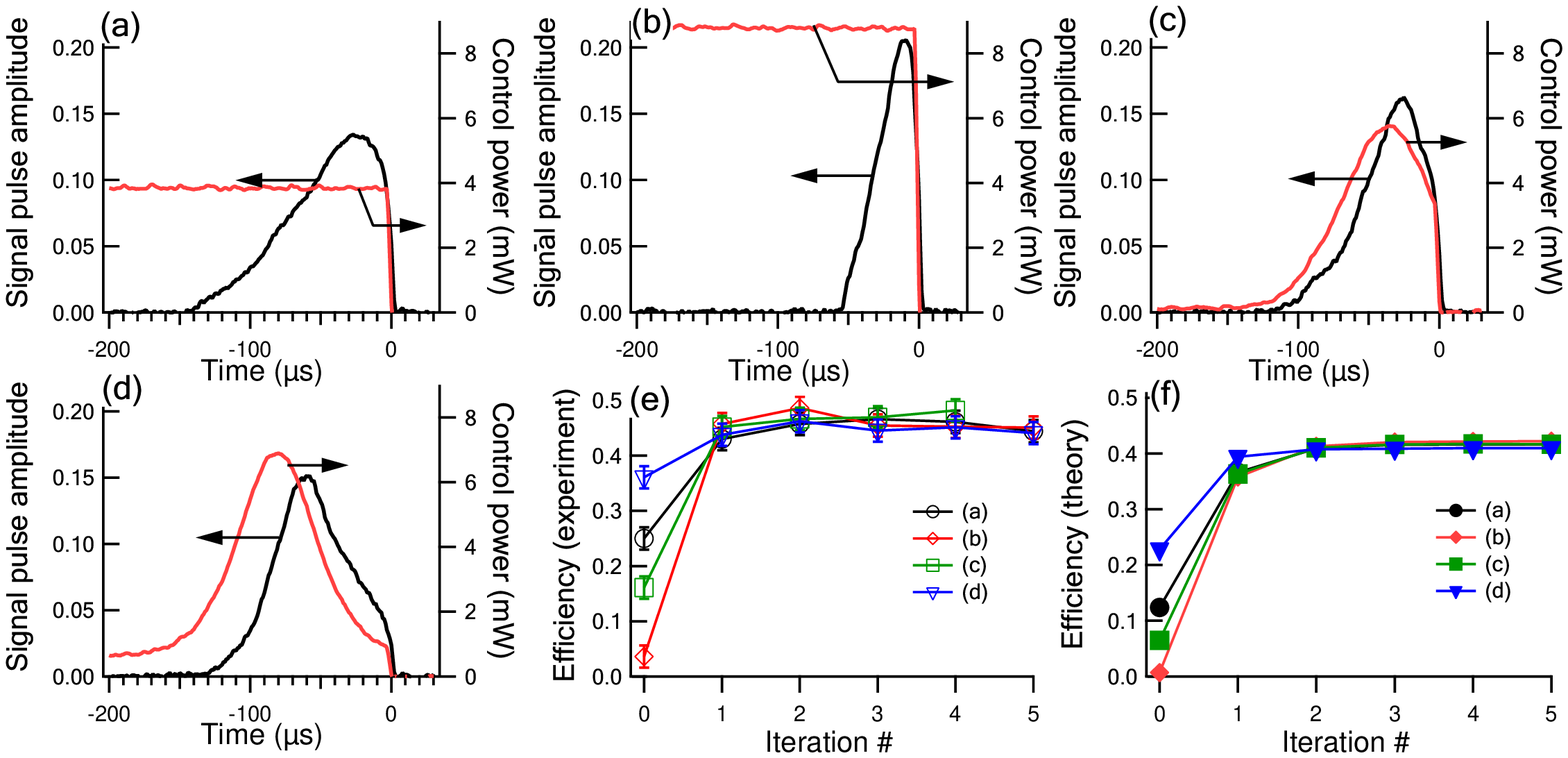}
\caption{(color online) (a)-(d) Examples of optimal input signal pulses
determined experimentally by application of the iterative optimization
procedure, for different control field profiles. In all cases, the initial
(trial) signal pulse has the same Gaussian pulse-shape and amplitude as the
trial pulse used in the data shown in Fig.~\ref{SampleProbePulses.fig}. (e)
Experimentally measured and (f) calculated write/store/retrieve efficiencies as
functions of the iteration number.} \label{VarContrPulses.fig}
\end{figure*}

\emph{3.} For a given optical depth, different control field profiles result in
different optimal signal pulse-shapes but yield the same maximum efficiency,
provided spin-coherence decay during the writing and retrieval stages is small. See
Fig.~\ref{VarContrPulses.fig}.

We performed these experimental demonstrations using a standard Rb vapor EIT
setup, similar to that described in Ref.~\cite{JMO}. A cylindrical
$7.5$~cm-long glass cell containing isotopically enriched ${}^{87}$Rb and
$40$~Torr Ne buffer gas was mounted inside a three-layer magnetic shield to
reduce stray magnetic fields. The Rb vapor cell was typically operated at a
temperature $\simeq60^\circ$C, corresponding to a Rb vapor density
$\simeq2.5\times 10^{11}~\mathrm{cm}^{-3}$ and an optical depth $d\simeq9.0$.
Optical fields near the Rb $D1$ transition (795 nm) were used for EIT and light
storage. These fields were created by phase-modulating the output of an
external-cavity diode laser using an electro-optical modulator (EOM) operating
at the ground state hyperfine frequency of ${}^{87}$Rb (6.8 GHz). The laser
carrier frequency was tuned to the $5{}^2\mathrm{S}_{1/2}F=2 \rightarrow
5{}^2\mathrm{P}_{1/2} F^\prime=2$ transition and served as the control field
during light storage; while the high-frequency modulation sideband, resonant
with the $5{}^2\mathrm{S}_{1/2}F=1 \rightarrow 5{}^2\mathrm{P}_{1/2}
F^\prime=2$ transition, served as the signal field. The amplitudes of the
control and signal fields could be changed independently by simultaneously
adjusting the EOM amplitude and the total intensity in the laser beam using an
acousto-optical modulator (AOM). Typical peak control field and signal pulse
powers $\sim 5$~mW and $0.1$~mW, respectively. The laser beam was collimated to
a Gaussian cylindrical beam of relatively large diameter ($\simeq 7$~mm) and
then circularly polarized using a quarter-wave plate before entering the vapor
cell. The Rb atom diffusion time out of the laser beam ($\simeq 7$~ms) was long
enough to have negligible effects~\cite{xiao}. Small, remnant magnetic fields
were the leading source of spin decoherence, with typical spin-wave decay time
constants $\simeq2$~ms. We used relatively short pulses and storage times, such
that spin decoherence had a negligible effect except for a modest reduction of
the efficiency of the storage process.

Fig.~\ref{SampleProbePulses.fig} shows an example implementation of the
iterative optimization procedure, using a step-like control field and a trial
input signal pulse with a Gaussian profile. Some portions of the first input
pulse were incoherently absorbed or escaped the cell before the control field
was turned off; but a fraction was successfully mapped into an atomic
spin-wave, stored for $400~\mu$s, and then retrieved and detected. This
retrieved signal pulse-shape was used to generate a time-reversed and
normalized input signal pulse for the next iteration. After a few iterations,
both the input and output signal pulses converged to fixed profiles, with the
write/store/retrieve efficiency  increasing with each iteration and reaching a
maximum. In general, different trial input pulses all converged to the same
optimal signal pulse-shape (e.g., see Fig.~\ref{VarSigPulses.fig}).  In
addition, systematic variation of the signal pulse-shape uniformly yielded
lower efficiencies than the pulse-shape given by the optimization procedure.

We performed similar optimization experiments for a wide range of control field
profiles. Some example results are shown in Fig.~\ref{VarContrPulses.fig}. In
general we found different optimized signal pulse-shapes for different control
field profiles; however, the optimized write/store/retrieve efficiency was
independent of the control field profile. This observation is consistent with
the theoretical prediction~\cite{gorshkov} that optimal light-storage
efficiency does not depend on the control field, but only on the optical depth,
provided spin decoherence and other loss mechanisms can be neglected during the writing and retrieval stages.

To compare our experimental results with theoretical calculations, we
approximated the 16-level structure of the ${}^{87}$Rb D$_1$ line with a single
three-level $\Lambda$-system and modeled stored light dynamics as
follows~\cite{fleischhauer,lukin03rmp,gorshkov}:
\begin{eqnarray} \label{eq1}
(\partial_t + c \partial_z) \e(z,t)&=&i g \sqrt{N} P(z,t),
\\
\partial_t P(z,t)&=&- \gamma P(z,t) + i g \sqrt{N} \e(z,t) + \nonumber \\
&& i \Omega(t-z/c) S(z,t),
\\ \label{eq3}
\partial_t S(z,t)&=&-\gamma_\textrm{s} S(z,t) + i \Omega(t-z/c) P(z,t).
\end{eqnarray}
Here $\e$ is the slowly-varying envelope for the signal field, $P$ is the transverse polarization of the atomic ensemble on the optical transition driven by the signal field, $S$ is the spin-wave envelope, and $\Omega$ is the control field Rabi frequency envelope. Also, $g\sqrt{N}$ is the coupling strength between the atomic
ensemble and the signal field, where $g$ is the corresponding one-photon Rabi frequency and $N$ is the number of atoms along the
laser beam that are available to participate in light storage after optical
pumping by the control field; $\gamma$ is the decoherence rate of $P$, due
primarily to buffer gas collisions; and $\gamma_\textrm{s}$ is the spin-wave
decoherence rate. To calculate $\Omega$, we used the dipole matrix element of
the $|F=2,m_F=1\rangle \rightarrow |F'=2,m_F=2\rangle$ transition, which was
the dominant control field transition for our experimental
conditions~\cite{Omeganote}. We approximated the laser beam profile as a
uniform cylindrical beam with the same diameter and total power as used in the
experiment.

We solved Eqs.~(\ref{eq1}-\ref{eq3}) analytically by adiabatically eliminating
$P$, an excellent approximation for our experimental
conditions. We then calculated the results of the iterative
optimization procedure: i.e., the output signal field (both signal leakage and
stored/retrieved pulses) as well as the generated input signal pulse for each
successive iteration. For these calculations we used values for $g\sqrt{N}$,
$\gamma$, $\gamma_\textrm{s}$, and the trial input signal pulse and control field appropriate for the particular experimental conditions. Example results of these calculations are shown in the
bottom panels of Fig.~\ref{SampleProbePulses.fig}. The calculated output signal
pulse-shapes are qualitatively similar to the experimental results and converge
to an optimal input signal pulse-shape within a few iteration steps. The
calculated efficiencies for the optimization procedure, shown in
Figs.~\ref{SampleProbePulses.fig} and ~\ref{VarContrPulses.fig}(f), are in
reasonable agreement with experiment. We also confirmed that the effects of
inhomogeneous Doppler broadening were small for the buffer gas pressure used in
our experiments, by repeating the calculations in a more realistic
approximation that included Doppler broadening of Rb atoms as well as velocity
changing collisions with buffer gas atoms.

In conclusion, we experimentally demonstrated an iterative optimization
procedure, based on time-reversal, to find the input signal pulse-shape that
maximizes the efficiency of light storage and retrieval. We confirmed the three
primary predictions of the theory underlying the optimization
procedure~\cite{gorshkov}: (i) efficiency grows with each iteration until the
input signal field converges to its optimal pulse-shape; (ii) the result of the
optimization procedure is independent of the initial (trial) signal
pulse-shape; and (iii) the optimal efficiency does not depend on the control
field temporal profile. We also performed theoretical calculations of the light
storage process and the optimization procedure, and found good qualitative
agreement with the experimental results, thus supporting  the interpretation
that optical depth is the key figure of merit for light storage efficiency. 
The optimization procedure should be applicable to both classical and quantum
signal pulses and to a wide range of ensemble systems. As one example, since
pulse-shape optimization with weak classical light pulses can be
straightforwardly performed, such optimization could be used to determine the
temporal profile of input quantum fields, for which mode-shape generation and
measurement are much more difficult to carry out. Also,
pulse-shape optimization 
of the kind demonstrated here in atomic ensembles could be applicable to other
systems, \emph{e.g.}, photonic crystals~\cite{Yanik04}.

We are grateful to M. Hohensee for useful discussions, and to M. Klein and Y.
Xiao for assistance in experiments. This work was supported by ONR, DARPA, NSF,
Danish Natural Science Research Council, Packard foundation, and the
Smithsonian Institution.


\begin{thebibliography}{99}
\frenchspacing

\bibitem{lukin03rmp}
M.~D.\  Lukin, Rev.\ Mod.\ Phys.\ \textbf{75}, 457 (2003).

\bibitem{julsgaard04}
B.\ Julsgaard \emph{et al.},
 Nature \textbf{432}, 482 (2004).

\bibitem{kraus06}
B. Kraus \emph{et al.}, 
Phys. Rev. A \textbf{73}, 020302 (2006).

\bibitem{harrispt}
S.\ E.\ Harris, Phys. Today \textbf{50}(7), 36 (1997).

\bibitem{fleischhauerrmp}
M.\  Fleischhauer, A.\ Imamoglu, and J.\ P.\ Marangos, Rev. Mod. Phys.
\textbf{77}, 633 (2005).

\bibitem{fleischhauer}
M.\ Fleischhauer and M.\ D.\ Lukin, Phys. Rev. Lett. \textbf{84}, 5094, (2000);
Phys. Rev. A \textbf{65}, 022314 (2002).

\bibitem{liu}
C.\ Liu, Z.\ Dutton, C.\ H.\ Behroozi, and L.\ V.\ Hau, Nature {\bf 409}, 490
(2001).

\bibitem{phillips}
D.\ F.\ Phillips \emph{et al.},
Phys. Rev. Lett. {\bf 86}, 783 (2001).

\bibitem{hemmer}
P.\ R.\ Hemmer  \emph{et al.}, Opt. Lett. {\bf 26}, 361 (2001); J.\ J.\
Longdell  \emph{et al.}, Phys. Rev. Lett. {\bf 95}, 063601 (2005).


\bibitem{mair}
A. Mair \emph{et al.},
Phys. Rev. A {\bf 65}, 031802(R) (2002).

\bibitem{kuz03etal}
A. Kuzmich \emph{et al.}, Nature {\bf 423}, 731 (2003).

\bibitem{vdwal03}
C. H. van der Wal \emph{et al.}, Science {\bf 301}, 196 (2003).

\bibitem{chou04}
C. W. Chou, S. V. Polyakov, A. Kuzmich, and H. J. Kimble, Phys. Rev. Lett. {\bf
92}, 213601 (2004).

\bibitem{matsukevich04}
D. N. Matsukevich and A. Kuzmich, Science {\bf 306}, 663 (2004).

\bibitem{eisaman04}
M.\ D.\ Eisaman \emph{et al.},
Phys. Rev. Lett. {\bf 93}, 233602 (2004).

\bibitem{matsukevich06}
D. N. Matsukevich \emph{et al.}, Phys. Rev. Lett. {\bf 97}, 013601 (2006).

\bibitem{chen06}
S. Chen \emph{et al.}, Phys. Rev. Lett. {\bf 97}, 173004 (2006).

\bibitem{chou05}
C. W. Chou \emph{et al.}, Nature \textbf{438}, 828 (2005).

\bibitem{kuzmich05}
T.\ Chaneli\`ere \emph{et al.},
Nature \textbf{438}, 833 (2005).

\bibitem{eisaman05}
M.\ D.\ Eisaman \emph{et al.},
Nature \textbf{438}, 837 (2005).

\bibitem{DLCZ} L.\ M.\ Duan, M.\ D.\ Lukin, J.\ I.\ Cirac, and P.\ Zoller, Nature \textbf{414}, 413 (2001).

\bibitem{kuzmich06}
T.\ Chaneli\`ere \emph{et al.},
Phys. Rev. Lett. {\bf 96}, 093604 (2006).

\bibitem{felinto06}
D.\ Felinto \emph{et al.}, Nature Phys. \textbf{2}, 844 (2006).

\bibitem{novikova05}
I.\ Novikova, M.\ Klein, D.\ F.\ Phillips, and R.\ L.\ Walsworth,
Proc.\ SPIE.\ \textbf{5735}, 87 (2005).

\bibitem{felinto05}
D.\ Felinto \emph{et al.}, Phys. Rev. A {\bf 72}, 053809 (2005).

\bibitem{laurat06}
J.\ Laurat \emph{et al.}, Opt. Expr. {\bf 14}, 6912 (2006).

\bibitem{thompson06}
J.\ K.\ Thompson, J.\ Simon, H.\ Loh, and V.\ Vuleti\'c, Science \textbf{313},
74 (2006).

\bibitem{gorshkov}
A.\ V.\ Gorshkov \emph{et al.}, quant-ph/0604037 (2006); quant-ph/0612082;
quant-ph/0612083; quant-ph/0612084.

\bibitem{dnote} We define the optical depth $d$ as describing $1/e$ amplitude attenuation of a weak resonant signal pulse with no control field.

\bibitem{JMO}
I.\ Novikova, Y.\ Xiao, D.\ F.\ Phillips, and R.\ L.\ Walsworth,
J.\ Mod.\ Opt.\ \textbf{52}, 2381 (2005).

\bibitem{xiao}
Y.\ Xiao, I.\ Novikova, D.\ F.\ Phillips, and R.\ L.\ Walsworth, Phys. Rev.
Lett. {\bf 96}, 043601 (2006).

\bibitem{Omeganote}
As in Refs.~\cite{fleischhauer,gorshkov}, we define the Rabi frequency $\Omega$
as $|\Omega|^2=\wp^2 I /(2 \hbar^2 \epsilon_0 c)$, where $\wp$ is the
transition dipole matrix element and $I$ is the control field intensity.

\bibitem{Yanik04} M.F. Yanik,
\textit{et.\ al.}, Phys.\ Rev.\ Lett. \textbf{93}, 233903 (2004).

\end{thebibliography}
\end{document}